# Detailed Structure of the Magnetic Excitation Spectra of YBa$_2$Cu$_3$O$_y$ and Its Implication on the Physical Characteristics of the Electron System

Masafumi Ito[1,2], Hiroshi Harashina[1,2], Yukio Yasui[1,2], Masaki Kanada[1,2], Satoshi Iikubo[1], ,Masatoshi SATO[1,2], Akito Kobayashi[1] and Kazuhisa Kakurai[2,3]*

[1]*Department of Physics, Division of Material Science, Nagoya University, Furo-cho, Chikusa-ku, Nagoya 464-8602*
[2]*CREST, Japan Science and Technology Corporation (JST)*
[3]*Neutron Scattering Laboratory, ISSP, The University of Tokyo, Shirakata 106-1, Tokai, Ibaraki 319-1195*

## Abstract

Detailed structure of the magnetic excitation spectra $\chi''(\boldsymbol{q},\omega)$ of the superconducting oxide YBa$_2$Cu$_3$O$_{6.5}$(the transition temperature $T_c \cong 52$ K) in the wave vector ($\boldsymbol{q}$)- and the energy($\omega$)-space, and its temperature ($T$) dependence have been studied. By adopting an effective energy dispersion of the quasi particles which can reproduce the shape of the Fermi surface and by introducing the exchange interaction between the Cu spins, a rather satisfactory agreement between the calculation and the experimentally observed data can be obtained. In the study, it has been found that the effects of the quasi particle-energy broadening on the excitation spectra $\chi''(\boldsymbol{q},\omega)$ is important. The sharp resonance peak observed at $\omega \sim 40$ meV for the optimally doped system of YBa$_2$Cu$_3$O$_y$ can be naturally reproduced by the present model.

corresponding author :M. Sato (e-mail address:e43247a@nucc.cc.nagoya-u.ac.jp)

*present address: Advance Science Research Center, JAERI, Tokai, Ibaraki 319-1195

§1. Introduction

Towards the complete description of the electronic behavior of high $T_c$ superconductors, it seems to be essential to understand their magnetic excitation spectra $\chi"(\boldsymbol{q},\omega)$, because the superconductivity itself is considered to originate from the magnetic interaction of their electrons. The behavior of the pseudo gap, which can be seen in $\chi"(\boldsymbol{q},\omega)$ and in the electronic tunneling density of states $N_S(\omega)$ even far above $T_c$ in the underdoped systems shows how the singlet correlation appears with decreasing temperature $T$.[1,2] The $\boldsymbol{q}$ and $\omega$-dependence of the magnetic excitation spectra (including the incommensurate→ commensurate→ incommensurate variation of the peak structure found in $YBa_2Cu_3O_y$(YBCO or YBCO$_y$) with increasing $\omega$ at low temperatures in the reciprocal space[3]), and the $T$-dependence of $\chi"(\boldsymbol{q},\omega)$, are expected to give us detailed information what kind of model is appropriate to treat the strongly correlated electrons of high $T_c$ systems.

The model of the "stripe order" has been proposed by Tranquada et al.[4] to explain the superlattice reflections at the incommensurate points in the reciprocal space of $La_{2-x-y}Nd_ySr_xCuO_4$ with x~1/8(1/8-doped LNSCO). If slowly fluctuating (dynamical) "stripes" exist in systems without having the long range order, structural characteristics of $\chi"(\boldsymbol{q},\omega)$ similar to those found in $\chi"(\boldsymbol{q},\omega=0)$ of $La_{2-x-y}Nd_ySr_xCuO_4$ may also be observed. The incommensurate peaks observed in superconducting $La_{2-x}Sr_xCuO_4$ and YBCO [5,6] systems at the $\boldsymbol{q}$ points similar to those in $\chi"(\boldsymbol{q},\omega=0)$ of 1/8-doped LNSCO might indicate that the expectation is correct.

If the "stripes", which can probably be considered to be primarily driven by the charge order commonly exist in high $T_c$ systems, introducing significant effects on the magnetic excitation spectra $\chi"(\boldsymbol{q},\omega)$, we may have to consider $\chi"(\boldsymbol{q},\omega)$ essentially in the charge ordered background, that is, we may have to treat their physical properties in this background. However, the problem does not seem to be so simple: Because the realistic pattern of the "stripes" is difficult to be determined, we cannot calculate the magnetic excitation spectra accurately enough to be compared with measured results. It is also important that as Kao et al.[7] have first pointed out, calculations which use an expression

$$\chi(\boldsymbol{q},\omega) = \chi^0(\boldsymbol{q},\omega)/\{1+J(\boldsymbol{q})\chi^0(\boldsymbol{q},\omega)\}, \qquad (1)$$

for the dynamical susceptibility, can also reproduce qualitatively the incommensurate→ commensurate→ incommensurate variation of $\chi"(\boldsymbol{q},\omega)$ stated above, where $\chi^0(\boldsymbol{q},\omega)$ is the Lindhard function of the electron system without the exchange coupling $J(\boldsymbol{q})$ between the Cu-Cu electrons. Moreover, it has been shown by the present authors' group[8] that the $\boldsymbol{q}$ and $\omega$-dependence of $\chi"(\boldsymbol{q},\omega)$ of $YBCO_{6.5}$ observed at 7 K can be reproduced well by the same expression as eq. (1), where $\chi^0(q,\omega)$ is calculated by adopting an effective energy($\varepsilon$) dispersion of the quasi particles and by considering their energy-broadening $\Gamma(\varepsilon)$. (As is stated later, the models in refs. 7 and 8 are appropriate to the electron system irrespective of whether it is the Fermi-liquid-like or not.[7,9,10]) These results indicate that the existence of the "stripes" is not necessarily required to explain the observed behavior of $\chi''(q,\omega)$, and those fluctuating slowly enough to affect the physical properties may not exist in the system

In the present work, we have measured the $T$-dependence of $\chi''(q,\omega)$ of YBCO$_{6.5}$ ($T_c \cong 52$ K) as well as its $q$- and $\omega$-dependence in detail, and found that the results can well be reproduced by the similar calculation to that stated above. We have also found that the sharp peak (resonance peak) which is well known to exist in the spectra at $\omega \sim 40$ meV can also be reproduced by the calculation. Based on these results, arguments if the slowly fluctuating "stripes" commonly exist or not are given.

## §2. Experiments

The magnetic excitation spectra were taken for various values of $\omega$ at 7 K, 57 K and 147 K with the neutron triple axis spectrometer ISSP-PONTA installed at JRR-3M of JAERI in Tokai on a single crystal of YBa$_2$Cu$_3$O$_{6.5}$ ($T_c \sim 52$ K), whose data at 7 K were also used in ref. 8. The crystal was oriented with the [110] and [001] axes in the scattering plane. The 002 reflection of Pyrolytic graphite (PG) was used both for the monochromator and analyzer. The horizontal collimations were 40'−40'−80'−80'. The effective vertical collimations were 80'−240'−480'−650', where the resolution width $\Delta q_V$ along the vertical direction (//[1$\bar{1}$0]) is mainly determined by the last two numbers typically as $\Delta q_V \cong 0.33$ and 0.46 Å$^{-1}$ for the scattered neutron energies $E_f$ =14.7 and 30.5 meV, respectively. Intensity profiles were taken by scanning $h$ of $Q$ =($h,h,l$) with the transfer energy $E$ being fixed at several values. For $E \leq 15$ meV $E_f$ was fixed at 14.7 meV and for 15 meV $\leq E \leq$ 24 meV it was fixed at 30.5 meV. Only for $E \leq 24$ meV, we can choose the spectrometer setting with $|l| \sim 2$, where the $q$-resolution width $\Delta q_H$ along the [110] direction is not too large to observe the incommensurate nature of $\chi''(q,\omega)$. The typical value of $\Delta q_H$ (=$\sqrt{2}\Delta h \times (2\pi/a)$, $a$ being the lattice constant) is 0.08 Å$^{-1}$. (The separation of the incommensurate peaks observed in the low energy region of $\chi''(q,\omega)$ of the present crystal along the scan direction, is ~0.19 Å$^{-1}$ in the reciprocal space.) For the higher $E$ region, we have to choose the value of $|l|$ ~5, at which the second peak of $\chi''(q,\omega)$ modulated along $|l|$ exists,[11,12] to satisfy the condition of the energy and momentum conservation. At the $Q$ point, $\Delta q_H$ becomes much larger and the peak separation of $\chi''(q,\omega)$ cannot be detected. The crystal was mounted in an Al can with He exchange gas which was attached to the cold finger of Displex type closed cycle refrigerator.

We add followings here. We estimated the $y$-value of the present crystal to be 6.5

from the lattice parameter c($\cong$11.738 Å). However, being judged from the observed $\chi''(\bm{q},\omega)$, the present crystal has larger hole-carrier concentration than that of the crystal labeled IMS No.29($T_c \cong$50 K), for which the oxygen number $y$ was also estimated to be ~6.5 and experimental data of $\chi''(\bm{q},\omega)$ were reported in the former papers.[13,14] The slightly higher $T_c$($\cong$52 K) than that of IMS No.29 also supports the presumption. Comparison of the present data with those of IMS No.29 will be presented in *3.2*.

## §3. Calculated and Experimental Results and their Comparison

### *3.1 Model of the calculation*

In the present model calculation of the magnetic excitation spectra $\chi''(\bm{q},\omega)$, we simply adopt eq.(1). As is emphasized in ref. 7, the expression is appropriate irrespective of whether the electron system can be treated as the Fermi liquid or not. Here, in order to properly calculate $\chi^0(\bm{q},\omega)$, we introduce the effective band parameters, $t_0$, $t_1=-t_0/6$ and $t_2=t_0/5$, which have the usual meanings and can reproduce the Fermi surface shape of YBa$_2$Cu$_3$O$_y$.[10,15] The intra atomic Coulomb interaction energy $U$ is set to be 0 (The choice gives essentially same results as those of ref. 7, because here the strong correlation effects are already included by adopting a small value of $t_0$( ~ -20 meV for the present crystal). The energy broadening $\Gamma(\varepsilon)$ is also introduced to consider effects of significant quasi particle damping. The band width obtained by the parameters $t_0$, $t_1$ and $t_2$ is similar to that obtained by the *d-p* model for the coherent in-gap band.[16] It seems to be also consistent with the width of the holon band of the *t-J* model.[10] For the exchange coupling, the nearest neighbor form of $J(\bm{q})= J(\cos q_x a+\cos q_y a)$ is used. The *d*-wave form of the energy gap $\Delta_s(\bm{k})=(\Delta_0/2)(\cos k_x a-\cos k_y a)$ is also used.

In the calculation, the absolute amplitudes of the superconducting gap of the underdoped system such as YBCO$_{6.5}$ and the pseudo gap above $T_c$ are taken to be equal and approximately $T$-independent in the present $T$ region, by considering that the pseudo gap persists up to very high temperature, while for the optimally doped YBCO system, the absolute amplitudes of the gap is considered to be $T$-dependent. As for the other parameters, the $J$ values have been chosen so that the commensurate peak of $\chi''(\bm{q},\omega)$ appears at the experimentally observed energy position and the chemical potential $\mu$ has been determined so that the $\bm{q}$ position of the incommensurate peak of $\chi''(\bm{q},\omega)$ at low temperatures and at relatively low energies agree with those observed for YBCO$y$.[6,17] The broadening $\Gamma(\varepsilon)$ of the quasi particle energy is simply assumed to be isotropic and a model which roughly reproduces characteristics of the experimentally observed $T$- and $\varepsilon$-dependence of $\Gamma$[18-21] is used.

In the numerical calculations of $\chi^0(\bm{q},\omega)$ and $\chi(\bm{q},\omega)$, the two-dimensional (2D) $q_x$-$q_y$ plane of the reciprocal space ($|q_x|, |q_y|<\pi/a$. Hereafter, the indices of $\bm{q}$, $x$ , $y$ and $z$ correspond to the $a$-, $b$- and $c$-axes of YBa$_2$Cu$_3$O$_y$ crystals, respectively.) is divided into 100×100 cells and the energy-integration is taken at the step of 0.5 meV, which is always smaller than the quasi particle broadening $\Gamma(\varepsilon)$, as in the case of ref. 8.

*3.2 Experimental and calculated results*

In Fig. 1 magnetic excitation spectra, $\chi''(\boldsymbol{q},\omega)$ calculated for the present YBCO$_{6.5}$ at 7 K are shown along $(h,1/2)$ in the 2D reciprocal space. They are obtained for the following parameters. $2\Delta_0$ =88 meV and $\mu$=0 meV. $\Gamma$(the full width at half maximum) is taken to be a small and $\varepsilon$-independent value of 4 meV and $J$=60.5 meV in the bottom figure, while in the top figure $\Gamma(\varepsilon)=\Gamma_0+(\Gamma_h-\Gamma_0)\times F(\varepsilon)$ is used to reproduce the characteristics of the $\Gamma$-$\varepsilon$ curve observed at low temperatures(see the inset of Fig. 1), where $F(\varepsilon)$ satisfies $F(0)=0$ and $F(2\Delta_0)=1$ and has a polynomial form up to the cubic order of $\varepsilon$, $\Gamma_0$ is $T$-dependent and has a value of 4 meV at 7 K and $\Gamma_h$ is a $T$-independent constant of 50 meV. $J$=58.5 meV. (For $2\Delta_0$, the same value as in the case of ref. 8 is used, because qualitative results are rather insensitive to the value and the other parameters are equal to the corresponding ones which can explain the experimental data, as shown later.)

One can see the incommensurate→ commensurate→ incommensurate variation with increasing $\omega$ in both figures. The commensurate peak is at the energy of ~28 meV, which is consistent with the experimental observation.[17,22] It should be noted here that the parameters $\Gamma(\varepsilon)$ and $J$ chosen in the present calculation are different from those used in the previous paper.[8] It is due to a following fact. Here, the quasi particle propagator is described as $G\sim1/(\omega-\varepsilon+i\Gamma)$ with $\Gamma$ being determined as a function of $\varepsilon$, as shown in the inset of Fig. 1, while in the previous paper, the broadening was given by $\Gamma(\omega)$ instead of $\Gamma(\varepsilon)$. If the absolute value of $\Gamma$ is not large, the difference does not change the spectra $\chi''(\boldsymbol{q},\omega)$ significantly. However, for $\Gamma$ as large as the one found for the present system, the parameters obtained by fitting the calculated results to the observed data changes significantly.

After convoluting the calculated $\chi''(\boldsymbol{q},\omega)$ with the resolution function, we obtain curves which can be directly fitted to the experimental data.(Detailed description of this resolution convolution can be found in the previous paper[8].) Here, $\Gamma_0$ has been set to be a small value of 4 meV. The parameters $\Gamma_h$=50 meV and $J$=58.5 meV have been found to give the satisfactory fitting, as shown in Fig. 2(a), to the data taken at 7 K by scanning the scattering vector $\boldsymbol{Q}$ along $(h,h,-2)$. It should be noted that here, a common scale factor is used to all the $\omega$ values. (The values of $\Gamma_h$ and $J$ have already been listed in describing the calculated data in Fig. 1.) For comparison, the results obtained for the same set of parameters as that for the bottom panel of Fig. 1 are shown in Fig. 2(b). As far as we use the $\varepsilon$-independent small $\Gamma$, it is difficult to explain the observed data even when the other parameters are changed. Then, it is clear that the consideration of the quasi particle broadening $\Gamma(\varepsilon)$ shown in the inset of Fig. 1, is important for the explanation of the experimentally observed $\chi''(\boldsymbol{q},\omega)$ of the YBCO system.

Figure 3(a) shows the results obtained for the present YBCO$_{6.5}$ at 57 K with the same parameters as those for Fig.2(a) except the value of $\Gamma_0$, which has been chosen here to be 10 meV. Again, the agreement between the experimental and the calculated results is rather satisfactory, though a slight disagreement can be observed at low energies. Figure 3(b) shows the results obtained for the same parameters as those to obtain the results shown in Fig. 2(b). Again, using the $\varepsilon$-independent small $\Gamma$, we cannot consistently explain the results(see the data at $\omega$=24 meV), even if the other parameters are changed.

Figure 4(a) shows the results obtained at 147 K with the same parameters as those for Figs. 2(a) and 3(a) except the value of $\Gamma_0$, which has been chosen to be 40 meV. The agreement between the experimental and the calculated results seems to be satisfactory. Figure 4(b) shows the results obtained for the same parameters as those corresponding to Figs. 2(b) and 3(b). For the $\varepsilon$-independent small $\Gamma$, the disagreement is getting more significant with increasing $T$.

We have found, in the above analyses that the use of $J$=58.5 meV can reproduce the observed $\chi''(\boldsymbol{q},\omega)$ well. The value is roughly consistent with the experimental value estimated by Johnston[23] for La$_{2-x}$Sr$_x$CuO$_4$ with $x$~0.1 and it seems to be also consistent with theoretically estimated value.[7,24] The $T$-dependence of $\Gamma_0$ found through the fittings at three temperature points seems to be reasonable, being judged by the data reported previously by other methods.[18-21] Figure 5 shows the $\omega$-dependence of the $\boldsymbol{q}$-dependent part of $\chi''(\boldsymbol{q},\omega)$ at (1/2,1/2) in the reciprocal space observed at 7 K and 147 K. The calculated curves are also shown by the solid and chain lines, respectively, by using the data shown in Figs. 2(a) and 4(a).

All these data indicate that the present model can explain the characteristics of $\chi''(\boldsymbol{q},\omega)$ rather well, if we take into account the broadening $\Gamma(\varepsilon)$ of the quasi particle energy. In this sense, we do not need the introduction of the dynamical "stripes". It is also noted that the broadening $\Gamma$ is important from the view point that it suppresses the tendency of the antiferromagnetic ordering: From the calculation, the nonzero $|\Delta_0|$ can be found to suppress the ordering, and it can be also found that, even when $\Delta_0$~0, the large broadening $\Gamma$ suppresses the antiferromagnetic ordering.

Now, we make brief comments on the differences between the present experimental data and those reported in the former papers[13,14] for the crystal IMS No.29, whose oxygen number was estimated to be ~6.5. In IMS No.29, gap-like structure (pseudo gap) was not observed in $\chi''(\boldsymbol{q},\omega)$. The incommensurate peak structure was not observed, either, and the data were analyzed by considering the antiferromagnetic correlation of the electron system. The absence of the gap-like structure in $\chi''(\boldsymbol{q},\omega)$ of IMS No.29 indicates that its hole-carrier number $p$ is smaller than that of the present one, because the magnitude of the experimental (pseudo) gap observed in $\chi"(\boldsymbol{q},\omega)$ increases with $p$ or $y$, or IMS No.29 can be considered to be located at a position closer to the boundary between the antiferromagnetic and superconducting phases than the present one. Then, for relatively small values of $\omega$, the $\boldsymbol{q}$-position of the peak of $\chi"(\boldsymbol{q},\omega)$ of IMS No.29 may be closer (even if it is at the incommensurate position) to the antiferromagnetic Bragg point $\boldsymbol{q}_{AF}=(\pi/a, \pi/a)$ than that of the present one, and therefore the resolution effects is expected to mask the incommensurate structure. The improvement of the method to prepare crystals with spatially homogeneous $y$ may also be an origin of the differences: The crystal IMS No.29 prepared in the rather early stage of the study may have larger width of the $y$ distribution, and therefore the contribution to the observed spectra in the relatively low energy region from parts of the sample near the phase boundary is enhanced, because the spectra is stronger at low energies for $y$ closer to the boundary, reflecting the strong antiferromagnetic spin correlation. One of other possible origins of the differences is, we think, an effect of the large energy broadening $\Gamma$ of the quasi particles near the phase boundary: As the system approaches the boundary, the superconducting transition temperature $T_c$ becomes small and the broadening $\Gamma_0$ in the $T$ region studied here does not become small as compared with $\Gamma_h$. For this large broadening, we just expect, from the calculation, the commensurate structure of $\chi"(\boldsymbol{q},\omega)$.

We have so far presented both the calculated and the experimental data of $YBCO_{6.5}$. Next, we discuss if the present model can reproduce the data of the optimally carrier-doped YBCO system. In particular, it is important to see if the sharp resonance peak found in the system[25,26] can be reproduced. For the calculation, we have tried to use a parameter set, $t_0$, $t_1$ and $t_2$ larger than those used for $YBCO_{6.5}$, with the ratios of the effective parameters being fixed not to change the Fermi surface shape, and for $t_0$=-25 meV (and $t_1$=-$t_0$/6 and $t_2$=$t_0$/5) and for the $\Gamma$-$\varepsilon$ curves A and B in the top right panel of Fig. 6 (For all the curves $\Gamma_0$=2 meV is used in the expression $\Gamma(\varepsilon)=\Gamma_0+(\Gamma_h-\Gamma_0)\times F(\varepsilon)$ and $\Gamma_h$ = 20 and 15 meV are used for the curves A and B, respectively.), we have obtained at 5 K the magnetic excitation spectra $\chi"(\boldsymbol{q},\omega)$ shown in Figs. 6(a) and 6(b), respectively, along $(h,1/2)$ in the 2D reciprocal space. Other parameters are as follows. $2\Delta_0$=70 meV, $\mu$=-5 meV, $J$=56 meV. We find that for rather small value of $\Gamma_h$, the sharp peak at ~40 meV is reproduced. It might indicate that the electrons of the system approach the ordinary Fermi liquid with increasing the oxygen number $y$. However, following things seem to be important. If we use the curve C, which has a different form of $F(\varepsilon)$ from that used for $YBCO_{6.5}$ and for the curves A and B, we can also reproduce the sharp resonance peak as shown in Fig. 6(c) even for $\Gamma_h$ as large as 40 meV. Because $\Gamma(\varepsilon)$ has been just assumed to be isotropic and to have a form which roughly reproduce the broadening observed along the [110] direction, there is no firm reason that the form of $\Gamma(\varepsilon)$ does not depend on the value of $y$. Then, to distinguish which case of A, B and C is appropriate to the electrons of the optimally doped system, we have tried to calculate the spectral function $\chi''(\boldsymbol{q},\omega)$ above $T_c$, where $2\Delta_0 \cong 0$ and therefore $\Gamma_0 = \Gamma_h$ are expected, and obtained the featureless and rather broad energy distribution of the spectral weight $\chi''$ only for the large values of $\Gamma_h = \Gamma_0 \geq 40$ meV. (For the cases A and B, large spectral weights are found in the low energy region, which seems to contradict the actual case.) From these facts, we think that the change of $F(\varepsilon)$ toward the form shown by the curve C is plausible, as $y$ increases from 6.5 to ~6.95.

## §4. Discussion and Summary

We have shown that the present model calculation, which uses the effective band parameters and introduces the exchange coupling $J$ between the Cu spins can well reproduce the experimentally observed characteristics of $\chi''(\boldsymbol{q},\omega)$ of underdoped YBCO. systems. These results suggest two things. First, the effective band we have used is rather narrow and may correspond to the coherent in-gap band of the $d$-$p$ model[16] or the Zhang-Rice singlet of the $t$-$J$ model.[10,27] At low temperatures, the superconducting long range order develops and the energy broadening or the damping of the quasi particles is small in the region of $|\varepsilon| < 2\Delta_0$ (For $\Gamma_0$ we have used the value of 4 meV, it is just due to a technical requirement in the calculation that the energy step of the summation should be smaller than $\Gamma(\varepsilon)$.), but rather large for $|\varepsilon| > 2\Delta_0$. As $T$ increases, the broadening in the energy region $|\varepsilon| < 2\Delta_0$ increases, which may present information how the band tends to loose the well-defined quasi particle character.

Second, the present model does not require the existence of the dynamical "stripes". Then, the question is whether the dynamical "stripes" can also present a satisfactory explanation of the observed data of $\chi''(\boldsymbol{q},\omega)$. To answer this seems not to be simple, because detailed pattern of the "stripes" is not known. If we consider the quasi static "stripe" pattern of holes and antiferromagnetically correlated Cu spins, significant spectral weight is expected at low energies around the corresponding incommensurate position, as the diffuse scattering contribution. If the singlet correlation is superposed on the antiferromagnetic one, the gap-like structure appears. The problem is what kind of $\omega$-dependence of the spectral weight is expected in such the system. The results of the model calculation reported by Batista et al.[28] seems to indicate that significant spectral weight remains down to the gap edge region, even if we consider the $d$-wave character of the gap. By their model, the energy of the resonance peak(or the commensurate peak) is roughly proportional to $\delta \times J$. Then, if we use the oxygen number($y$) dependence of $\delta$ reported by Dai et al.[15] that $\delta$ is almost constant for $y \geq 6.6$, it can hardly be expected that the resonance energy increases from ~28 meV to more than 40 meV when $y$ increases from 6.5 to 6.92, because $J$ is considered not to increase with the doped hole density.[23,24] In contrast with the result of the above consideration, the present calculation indicates that the increase of the effective band width with $y$ induces the rapid increase of the energy of the resonance peak without the increase of $J$, suggesting that the "stripes" may not have significant effects on the observed behavior of $\chi''(\boldsymbol{q},\omega)$. At least, the existence of slowly fluctuating "stripes" which affect the low energy physical properties of YBCO at low temperatures seems to be unlikely.

In summary, the magnetic excitation spectra $\chi''(\boldsymbol{q},\omega)$ observed for YBa$_2$Cu$_3$O$_y$ ($y \cong 6.5$) have been compared with the calculated results. In the calculation, the effective band parameters of the quasi particles are used and the exchange coupling $J$ between the Cu spins are considered. The consideration of the energy broadening of the quasi particles is shown to be important for the explanation of the spectra.

Figure captions

Fig. 1   Magnetic excitation spectra $\chi''(\mathbf{q},\omega)$ calculated for YBCO$_{6.5}$ at 7 K by eq. 1 for the effective band parameters $t_0$(=-20 meV), $t_1$=-$t_0$/6 and $t_2$=$t_0$/5 and $U$=0, are shown along the [10] direction in the two dimensional reciprocal space for several $\omega$ values. The bottom figure is for the small and $\varepsilon$-independent $\Gamma$ (=4 meV) and $J$=60.5 meV. The top figure is for $\Gamma(\varepsilon)$ shown in the inset. $J$=58.5 meV. See text for details.

Fig. 2   Neutron scattering intensities of the magnetic excitations taken at 7 K for a single crystal of YBa$_2$Cu$_3$O$_{6.5}$ by scanning along ($h,h,\sim$-2) at several fixed $E$ values (solid circles), where $E_f$=14.7 meV for $E \leq 15$ meV and, $E_f$=30.5 meV for $E$ >15 meV. Profiles obtained by convoluting the resolution functions with the calculated $\chi''(\mathbf{q},\omega)$ shown in the top and bottom panels of Fig. 1 are also shown by the solid lines in (a) and (b), respectively. Measurements at $E$=15 meV were carried out for both $E_f$ values of 14.7 and 30.5 meV and the scale factors at these energies are properly determined by adjusting the calculated profiles to the corresponding experimental ones at $E$ =15 meV. The zeros of the vertical axis are shifted upwards by 100, 200, 300, 200, 300, 500, 700 counts/14400kmon. for $E$ =12.5, 13.5, 15, 17.5, 20, 22.5 and 24 meV, respectively.

Fig. 3   In (a) and (b), results at 57 K are shown similarly to the case of Figs. 2(a) and 2(b), respectively. In calculating the curves in (a), $\Gamma_0$=10 meV is used and the other parameters except $T$ are chosen to be equal to those used in Fig. 2(a). In (b), all parameters used in the calculations except $T$ are equal to those used in the case of Fig. 2(b).

Fig. 4   In (a) and (b), results at 147 K are shown similarly to the case of Figs. 2(a) and 2(b), respectively. In calculating the curves in (a), $\Gamma_0$=40 meV is used and the other parameters except $T$ are chosen to be equal to those used in Figs. 2(a) and 3(a). In (b), all parameters used in the calculations except $T$ are equal to those used in the case of Figs. 2(b) and 3(b).

Fig. 5   Experimental and calculated peak intensities of the magnetic scattering at $\mathbf{Q}$=(1/2,1/2) in the two dimensional reciprocal space are shown at 7 K and 147 K, where the $\mathbf{Q}$-independent part of the intensities is removed.

Fig. 6   Magnetic excitation spectra $\chi''(\mathbf{q},\omega)$ calculated for the $\Gamma$-$\varepsilon$ curves in the top right panel (before the resolution convolution). The curves A, B and C correspond to Figs. (A), (B) and (C), respectively.

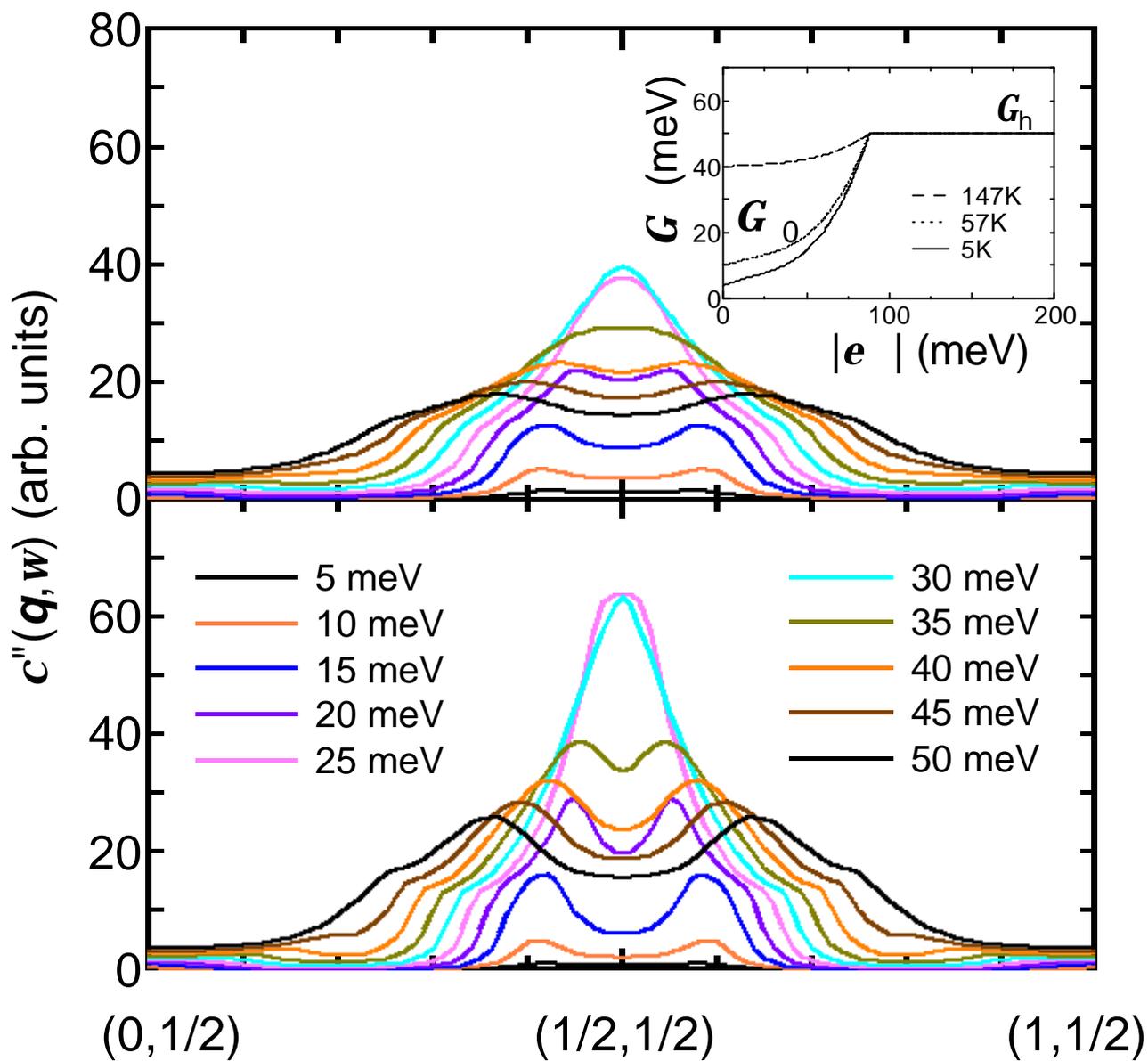

Figure 1

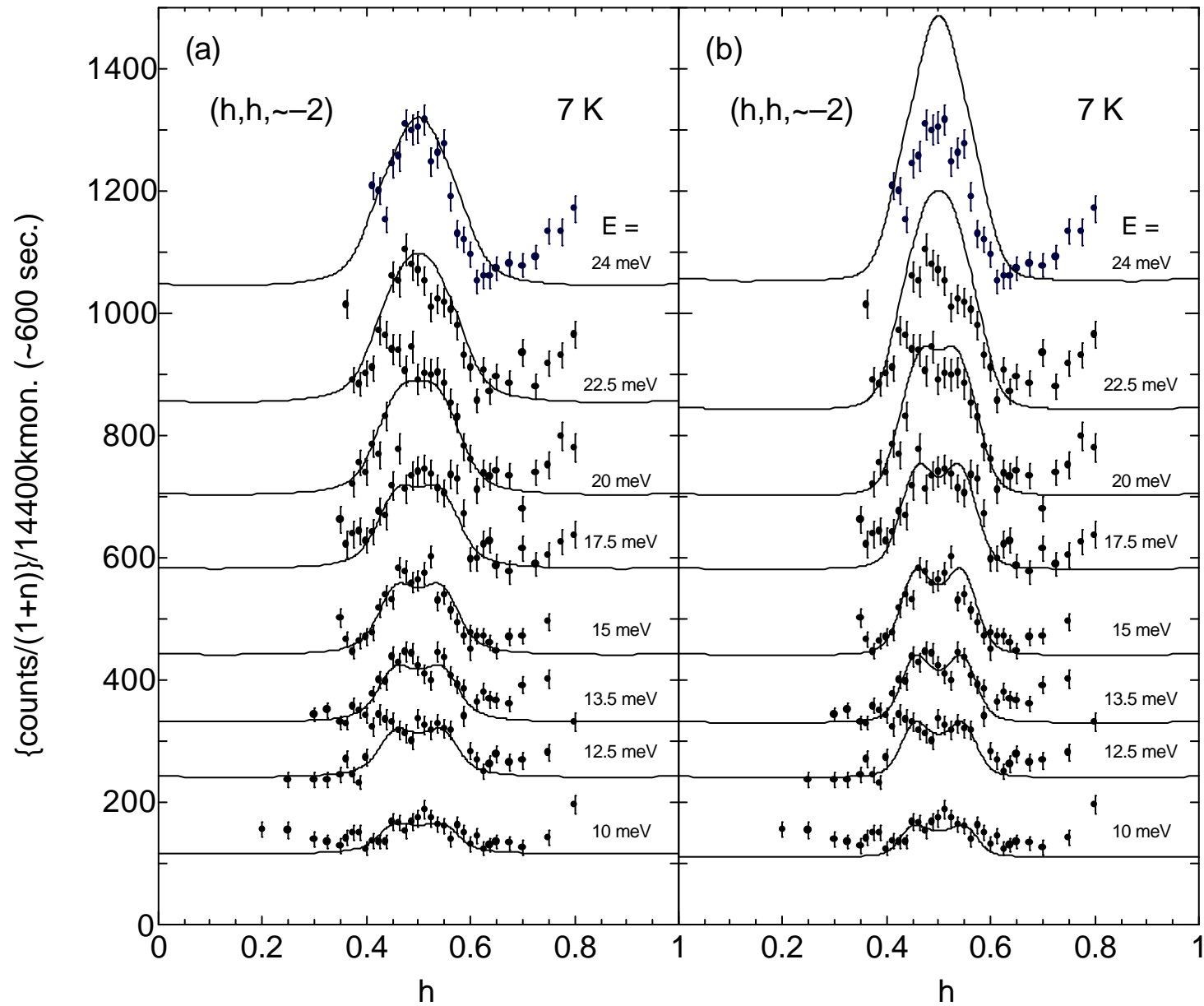

Figure 2

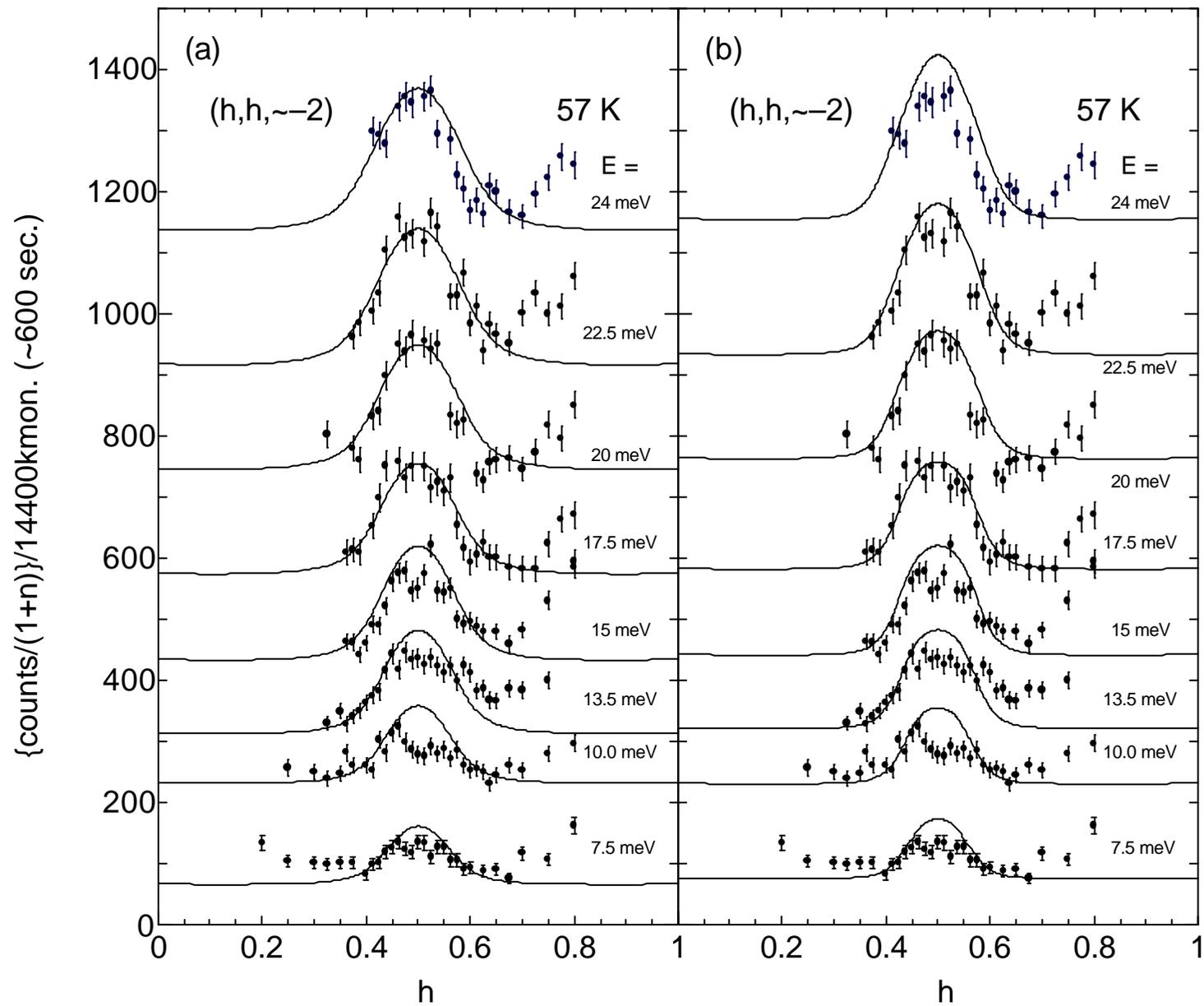

Figure 3

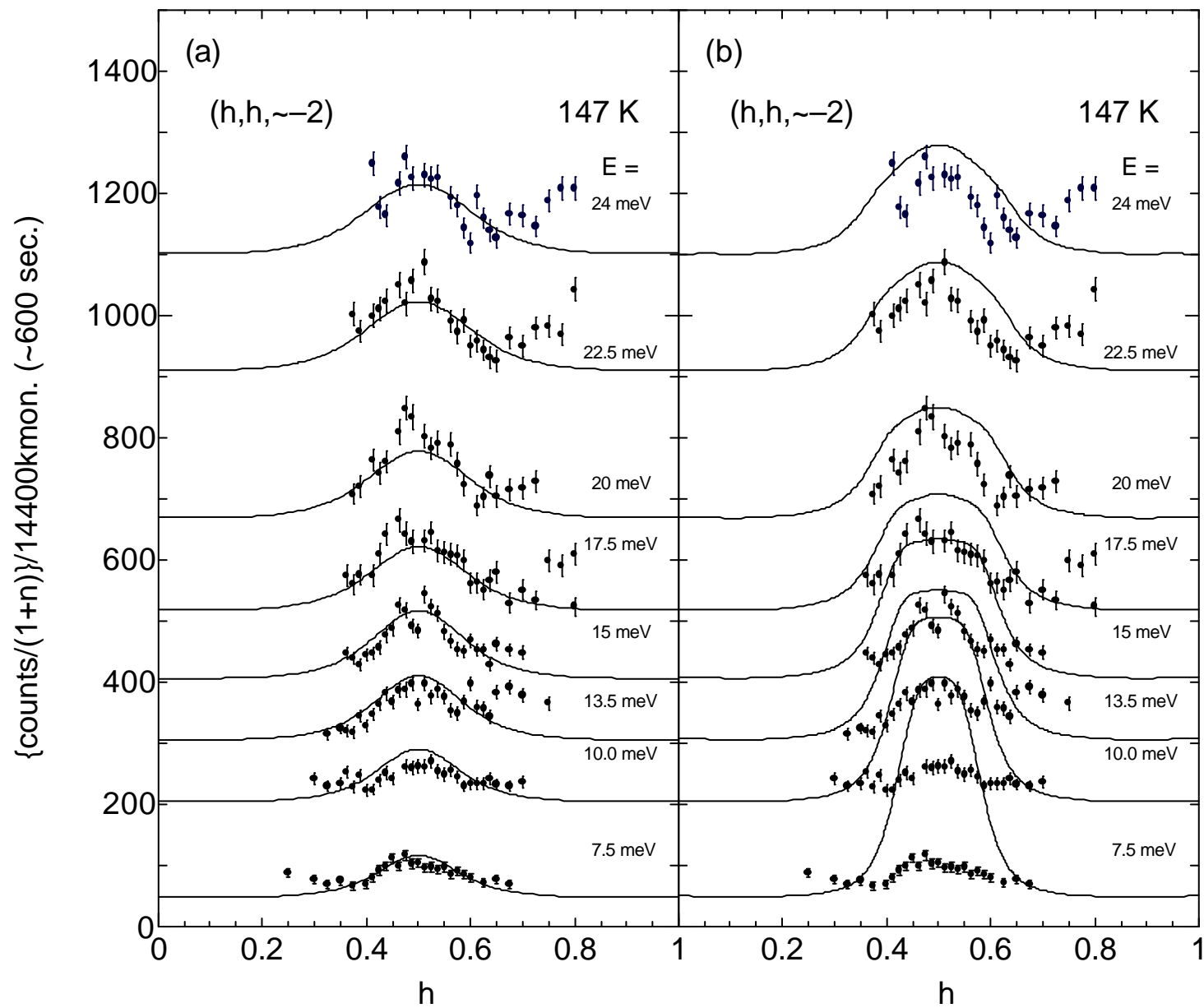

Figure 4

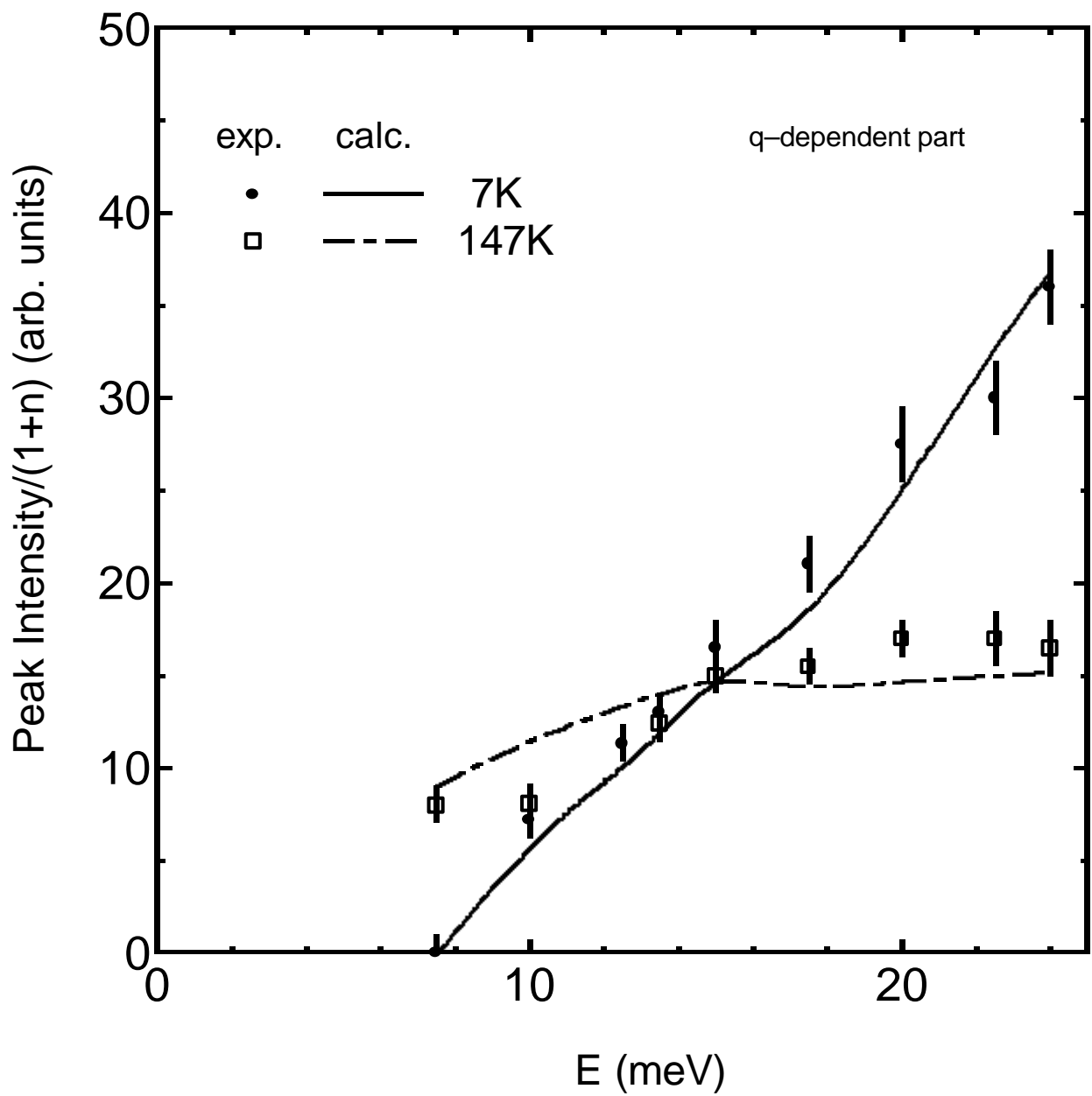

Figure 5

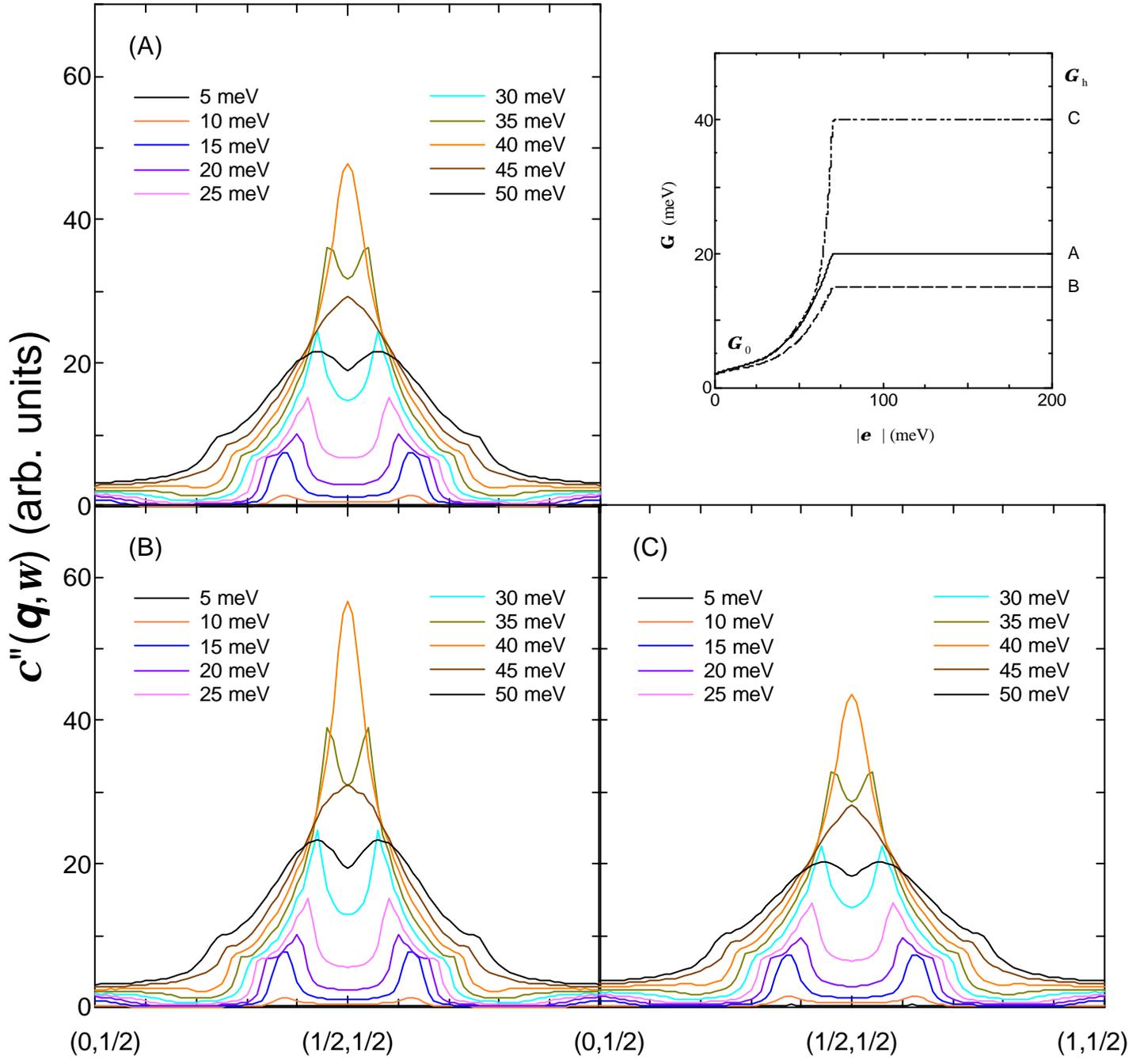

Figure 6